\documentclass[conference]{IEEEtran}
\IEEEoverridecommandlockouts
\usepackage[bottom=1.02in,top=0.71in,left=0.64in,right=0.64in]{geometry}
\usepackage{amsmath}
\usepackage{amsfonts}
\usepackage{amssymb}
\usepackage[ruled,vlined]{algorithm2e}
\usepackage{caption}
\usepackage{mathtools}  
\usepackage{amsthm}
\usepackage[]{algorithm2e}
\usepackage{booktabs}
\usepackage{subfigure}
\usepackage{tabulary}
\usepackage{footnote}
\usepackage[export]{adjustbox}
\usepackage{booktabs}
\usepackage[justification=centering]{caption}
\usepackage{comment}

\usepackage{cases}
\usepackage{graphicx}
\usepackage{cite}
\usepackage{multicol}
\usepackage{xcolor}
\usepackage{graphicx}
\graphicspath{{./}{figures/}}
\newcommand\numeq[1]%
  {\stackrel{\scriptscriptstyle(\mkern-1.5mu#1\mkern-1.5mu)}{=}}
\usepackage{stfloats}
\usepackage{cuted}
\usepackage{lipsum}  

\usepackage{dsfont}
\usepackage{amsmath,amssymb}

\newtheorem{Prob}{\textbf{Problem}}

\usepackage{optidef}
\usepackage{amsmath}
\title{Latency-aware End-to-end Multi-path Data Transmission for URLLC Services}

\author{\IEEEauthorblockN{Liu Cao$^*$, Abbas Kiani$^\dagger$, Amanda Xiang$^\dagger$, Kaippallimalil John$^\dagger$, Tony Saboorian$^\dagger$}
\IEEEauthorblockA{{$^*$Department of Electrical and Computer Engineering, University of Washington, Seattle, WA, USA} \\{$^\dagger$Wireless Research and Standards, Futurewei Technologies Inc., Addison, TX, USA}\\
Emails: liucao@uw.edu, \{abbas.kiani, amanda.xiang, john.kaippallimalil, tony.saboorian\}@futurewei.com}

}

\begin{document}

\maketitle
\thispagestyle{empty}
\begin{abstract}
5th Generation Mobile Communication Technology (5G) utilizes the Access Traffic Steering, Switching, and Splitting (ATSSS) rule to enable multi-path data transmission, which is currently being standardized. Recently, the 3rd Generation Partnership Project (3GPP) SA1 and SA2 have been working on the multi-path solution for possible improvement from different perspectives. However, the existing 3GPP multi-path solution has some limitations on ultra-reliable low-latency communication (URLLC) traffic in terms of reliability and latency requirements. In order to capture the potential gains of multi-path architecture in the context of URLLC services, this paper proposes a novel traffic splitting technique that can more efficiently enjoy the benefit of multi-path architecture in reducing user equipment (UE) uplink (UL) end-to-end (E2E) latency. In particular, we formulate an optimization framework that minimizes user's UL E2E latency via the joint optimization on the ratio of traffic assigned to each path and their corresponding transmit power. The performance of the proposed scheme is evaluated via well-designed simulations.

\end{abstract}


\section{Introduction}
\label{sec:intro}
 \IEEEPARstart{N}{owadays} a mobile device may establish multiple data paths simultaneously with the same application using different access technologies (e.g., 5G, 4G, WiFi, Satellite, etc.). Even connecting through different networks (e.g., private network, public network) to improve user quality of service (QoS) is becoming one of the feasible and desirable deployment options in beyond 5G (B5G). 3rd Generation Partnership
Project (3GPP) SA1 and SA2 have been working on end-to-end (E2E) multi-path solution since Release 16 \cite{3gpp.23.725} and the work will be continued on Release 18 and 19. Specifically, redundant transmission for high-reliability communication was first introduced \cite{3gpp.23.501}. The Phase 3 of further enhancing Access Traffic Steering, Switching and Splitting (ATSSS) has started in Release 18 \cite{3gpp.23.700}. More recently, Release 19 studied the upper layer ATSSS over dual 3GPP access\cite{3gpp.22.841}.

Although 3GPP has paid particular attention to the E2E multi-path solution for data transmission, it still has some limitations in the following aspects: 1) The current solution only takes into consideration between 3GPP access and non-3GPP access. However, Release 19 has planned to consider the ATSSS between two 3GPP access; 2) It does not satisfy the latency and the reliability requirement of URLLC traffic because all traffic is splitted and steered based on a static policy; 3) It lacks holistic and intelligent view on E2E QoS assurance, e.g., UE and network function (NF) make individual decisions based on simple inputs and relative static policy. 

The recent research literature has focused on the multi-path work to fulfill the latency and reliability requirements of data transmission \cite{simon2020atsc, ko2018joint, singh2016proportional, ba2022multiservice, hurtig2018low, yin2022routing, yin2021scheduling}. In particular, the authors in \cite{ba2022multiservice} investigated an optimization problem for traffic scheduling in NR and WLAN aggregation (NWA) which is mainly focused on enhanced mobile broadband (eMBB) services and does not take into consideration of specific requirements of URLLC services. The literature \cite{hurtig2018low} studied a multi-path transmission control protocol (MPTCP) scheduler, providing a good user experience for latency-sensitive applications when interface quality is asymmetric. However, this work considers the multi-path only from the transport layer perspective. In addition, machine learning (ML) based algorithms have been applied in the multi-path architecture. A joint power control and channel allocation scheme was developed in \cite{zhao2019joint} to reduce interference adaptively based on combining a reinforcement learning (RL) algorithm from the radio link side rather than the E2E side.

Motivated by the aforementioned issues, this paper proposes a novel E2E multi-path solution for URLLC data transmission. The major contributions of this paper are summarized as follows. 
\begin{itemize}
\item  We propose an optimization framework that minimizes the latency of user equipment (UE) uplink (UL) URLLC traffics by jointly optimizing the traffic ratio of each path and associated transmit power.
\item The proposed optimization framework is a dynamic policy because the decision in each time interval is determined by the combination of particular network conditions, traffic characteristics, and latency requirements. 
\item We compare the proposed multi-path solution with potential baselines under different scenarios, showing the advantages of the proposed multi-path solution.

\end{itemize}

The rest of this paper is organized as follows: Sec \ref{sec:sys_arc}  introduces the system architecture. The system model is illustrated in Sec \ref{sec:sys_model}. In Sec \ref{sec:sim}, we compare the multi-path solution with proposed baseline solutions in terms of the instant latency and the average latency via the simulation. Finally, section \ref{sec:con} draws the conclusions for this paper. 




\section{System Architecture}
\label{sec:sys_arc}

\begin{figure}[t]
    \centering
    \includegraphics[width=.35\textwidth]{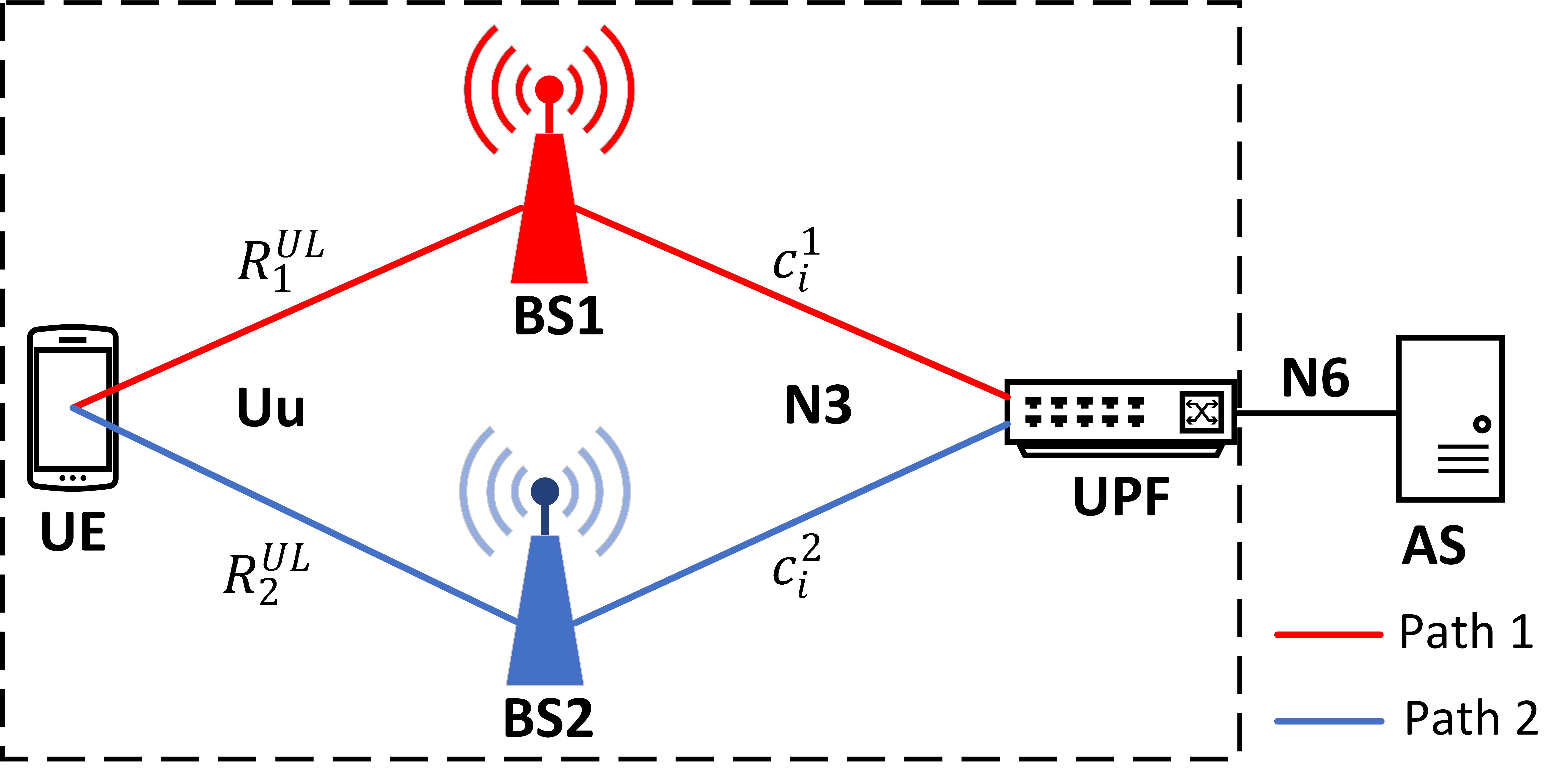}
    \caption{E2E multi-path system architecture.}
    \label{fig:E2Esysarch}
\end{figure}
As Fig. \ref{fig:E2Esysarch} shows, we consider an E2E system architecture where a single UE that has multiple traffic types for UL transmission connects to an application server (AS) via radio access network (RAN) and core network (CN) \cite{kaippallimalil20205g}. UE connects to each base station (BS) via a wireless link (i.e., Uu interface), and each BS is connected to the user plane function (UPF) via an N3 link. The UPF then connects to the AS via an N6 link. We assume that the UE is always in the coverage overlapping area of two BSs by constantly walking between them. As a result, the UE always has two paths that are established between the UE and UPF via BS1 and BS2, respectively. We consider an optimization framework where UE makes decisions for all UL traffic types by considering the joint optimization on the traffic ratio allocated to two paths and the associated transmit power\footnote{Since all decisions are made at the UE side, we are not optimizing any of BS resources such as the bandwidth assigned by the BS.}. To formulate such an optimization framework, we consider a time-slotted system where each time interval has a fixed duration, $t_s$ \footnote{The time-slotted model does not capture the real-time changes, however, the shorter the duration of a time interval is, the more accurate the real-time changes can be captured. Meanwhile, the duration of a time interval should be also large enough to avoid too many computing tasks on the UE side.}, and UE makes optimization decisions at the beginning of each time interval, e.g., $t_1, t_1 + t_s, t_1 + 2t_s, \cdots$.  It is assumed that the optimization decision, which is determined at the beginning of this time interval, will last for the entire time interval. Notice that different traffic types should have different transmission decisions made by UE within a time interval due to different traffic characteristics.

We denote $R_1^{UL}$ (bits/sec) and $R_2^{UL}$ (bits/sec) be the UL data rate of the wireless link at the beginning of each time interval on path 1 and path 2, respectively. Denote the QoS flow data rate of traffic type $i\in I$ over N3 links between BS and UPF on two paths as $c_i^1$ and $c_i^2$ (bits/sec), respectively. Note that $R_1^{UL}$, $R_2^{UL}$, $c_i^1$ and $c_i^2$ can be obtained at the UE side through the feedback from the BSs. The data rate of radio links ($R_1^{UL}$ or $R_2^{UL}$) is always assumed to be the bottleneck when compared with the data rate of QoS flow ($c_i^1$ or $c_i^2$). Therefore, the impact of queuing at the BSs on the UL optimization decision is almost negligible \footnote{This paper considers the queuing packets at the UE at the beginning of each time interval; however, it is assumed the queuing delay at BS is much lower than the transmission latency for the UL data transmission.}.

\section{System Model}
\label{sec:sys_model}
In this section, we provide a system model for the proposed E2E multi-path solution. Let $0 \le \alpha_i \le 1$, $\alpha_i\in\mathbb{R}$ be the portion of traffic type $i\in I, |I| = N_t$ routed to path 1 and $1-\alpha_i$ be the portion of traffic type $i$ routed to path 2. As mentioned in Section \ref{sec:sys_arc}, at the beginning of each time interval, the UE optimization solver decides the traffic ratio $\alpha_i$ for traffic type $i$ and the traffic ratio $1-\alpha_i$ for traffic type $i\in I$ on path 1 and path 2, respectively. 

According to Fig. \ref{fig:E2Esysarch}, the UL E2E latency on each path includes the latency over the wireless link and the latency over the N3 link. Let's assume the number of arrival packets of traffic type $i$ is random variable $\lambda_i^{UE}$ which follows a Poisson distribution with the given mean $E[\lambda_i^{UE}] = \overline{\lambda_i^{UE}}$. The packet size of traffic type $i$ is $M_i$ (number of bits) where the packet sizes are assumed to be identical for each traffic type. Meanwhile, we denote the number of packets of traffic type $i$ at UE queue at the beginning of each time interval as $n_i^q$, which also follows a Poisson distribution with the given mean $E[n_i^q] = \overline{n_i^q}$. Therefore, the estimated instant latency of traffic type $i$ over the wireless link on path 1 at the beginning of a time interval is given by
{\small \begin{align}\label{eqn: UulatencyPath1}
T_i^1(\alpha_i, P_1) = \frac{\alpha_i(\lambda_i^{UE}+n_i^q)M_i}{R_1^{UL}(P_1)},  
\end{align}}
where $R_1^{UL}(P_1)$ is the data rate of wireless link on path 1 at the beginning of the time interval, which is a function of transmit power $P_1$ allocated to path 1. $R_1^{UL}(P_1)$ can be expressed as
{\small\begin{align}\label{eqn: R1}
R_1^{UL}(P_1) = B_1\mathrm{log_2}\left(1+\frac{P_{RX}(P_1)}{N}\right),
\end{align}}
where $B_1$ is the bandwidth assigned by BS 1. $N$ is the average noise power, and $P_{RX}(P_1)$ is the receiver (RX) power at the BS 1, given that the allocated transmit power from UE is $P_1$ \footnote{The computation of $P_{RX}(P_1)$ follows the rule of large scale fading, which considers both the pathloss and shadowing effects.}. 

Meanwhile, the estimated instant latency of traffic type $i$ over N3 link on path 1 is 
{\small\begin{align}\label{eqn: N3latencyPath1}
D_i^1(\alpha_i) = \frac{\alpha_i(\lambda_i^{UE}+n_i^q)M_i}{c_i^1},  
\end{align}}
where $c_i^1$ (bits/sec) is the guaranteed bit rate (GBR) of traffic type $i$ over N3 link on path 1, which is assumed to follow a uniform distribution characterized by a lower bound and a upper bound. As a result, the estimated total instant UL latency of transmitting data for traffic type $i\in I$ via path 1 is given by
{\small\begin{align}\label{eqn: latencyPath1}
u_i^1(\alpha_i, P_1) = T_i^1(\alpha_i, P_1) + D_i^1(\alpha_i).  
\end{align}}
Similarly, the estimated total instant UL latency of transmitting data for traffic type $i\in I$ via path 2 is 
{\small\begin{align}\label{eqn: latencyPath2}
u_i^2(\alpha_i, P_2) = T_i^2(\alpha_i, P_2) + D_i^2(\alpha_i),
\end{align}}
where $T_i^2(\alpha_i, P_2)$ is the instant latency of traffic type $i$ over the wireless link on path 2 at the beginning of a time interval which is expressed as
{\small\begin{align}\label{eqn: UulatencyPath2}
T_i^2(\alpha_i, P_2) = \frac{(1-\alpha_i)(\lambda_i^{UE}+n_i^q)M_i}{R_2^{UL}(P_2)},  
\end{align}}
where $R_2^{UL}(P_2)$ is the data rate of the wireless link on path 2 at the beginning of a time interval, which is a function of transmit power $P_2$ allocated to path 2. Then $R_2^{UL}(P_2)$ is expressed as
{\small\begin{align}\label{eqn: R2}
R_2^{UL}(P_2) = B_2\mathrm{log_2}\left(1+\frac{P_{RX}(P_2)}{N}\right),
\end{align}}
where $B_2$ is the bandwidth assigned by BS 2, and $P_{RX}(P_2)$ is the RX power at the BS 2, given that the allocated transmit power from UE is $P_2$. Meanwhile, the estimated instant latency of traffic type $i$ over the N3 link on path 2 is 
{\small\begin{align}\label{eqn: N3latencyPath2}
D_i^2(\alpha_i) = \frac{(1-\alpha_i)(\lambda_i^{UE}+n_i^q)M_i}{c_i^2},  
\end{align}}
where $c_i^2$ (bits/sec) is the GBR of traffic type $i$ over N3 link on path 2. Since our objective is to minimize the UE's latency of all traffic types which share the same frequency band, the optimization problem for the proposed E2E multi-path solution is proposed to be formulated as below:
{\small\begin{Prob}[Proposed UL E2E Multi-path Solution]
\label{prob:e2eMultipath}
\begin{equation}
\begin{aligned}
&\underset{\alpha_i, P_1, P_2}{\mathrm{argmin}} \quad \sum_{i=1}^{N_t} \mathrm{max}\left( u_i^1(\alpha_i, P_1), u_i^2(\alpha_i, P_2)\right)  \\
\mathrm{s.t.} ~~~~&\mathrm{C1}: 0\leq u_i^1(\alpha_i, P_1) \leq T_i^{max}, \quad i \in I, \\
&\mathrm{C2}: 0\leq u_i^2(\alpha_i, P_2) \leq T_i^{max}, \quad i \in I, \\
&\mathrm{C3}: 0 \leq \alpha_i \leq 1, \quad i \in I,\\
& \mathrm{C4}: 0 \leq P_1 \leq P_{tot},  \\
& \mathrm{C5}: 0 \leq P_2 \leq P_{tot},\\
& \mathrm{C6}: P_1 + P_2 =  P_{tot},\\
\label{p:mian}
\end{aligned}
\end{equation}
\end{Prob}}
where $T_i^{max}$ is the maximum latency constraint of traffic type $i$. $N_t$ is the number of traffic types. $u_i^1(\alpha_i, P_1)$ and $u_i^2(\alpha_i, P_2)$ are expressed in Eq. (\ref{eqn: latencyPath1}) and Eq. (\ref{eqn: latencyPath2}), respectively. As the instant latency of a traffic type in each time interval is determined by the maximum one between two paths, we thereby use $\mathrm{max}\left( u_i^1(\alpha_i, P_1), u_i^2(\alpha_i, P_2)\right)$ to represent the estimated instant latency of traffic type $i$. The constraints C1 and C2 indicate that the estimated instant latency of traffic type $i$ on each path should not exceed the maximum latency constraint $T_i^{max}$. C3 defines the range of traffic ratio on path 1. It should be noted that $\alpha_i = 0$ is the case when entire UL traffic goes via path 2, while $\alpha_i = 1$ is the case when entire UL traffic goes via path 1. The constraints C4 and C5 define the range of the transmit power allocated to each path, and C6 indicates that the sum of allocated transmit power should be the total transmit power $P_{tot}$. Since the optimization decision is made at the UE side, UE only optimizes the UE's transmit power and traffic ratio of each traffic type. Hence, optimization on the bandwidth assigned by the BS side is not considered in the solution.  
\section{Simulation}
\label{sec:sim}

In this section, we test the proposed UL E2E multi-path solution by conducting various simulations in Matlab. We propose two baseline solutions to justify the effectiveness of our proposed solution, which are described as follows:
\begin{itemize}
\item \textbf{Single-path solution}: UE uses a fixed path (i.e., path 1 or path 2) for all traffic types in all time intervals;
\item \textbf{Path-selection solution}: UE dynamically selects a better path between two paths for all traffic types in each time interval. 
\end{itemize}
In our simulation scenario, UE does a 2-D random walk between the two fixed BSs, and UE is always in their coverage overlapping areas. To investigate the performance difference between the multi-path and baselines, we consider the following more detailed scenarios:
\begin{itemize}
\item \textbf{Scenario 1}: UE walks around the center of two BSs;
\item \textbf{Scenario 2}: UE walks around one of BSs (e.g., BS2) in most time.
\end{itemize}

\begin{figure*}[t]
\begin{minipage}[t]{0.24\linewidth}
\centering
 \subfigure[Instant latency CDF of traffic X.]
{\includegraphics[width=1.7in]{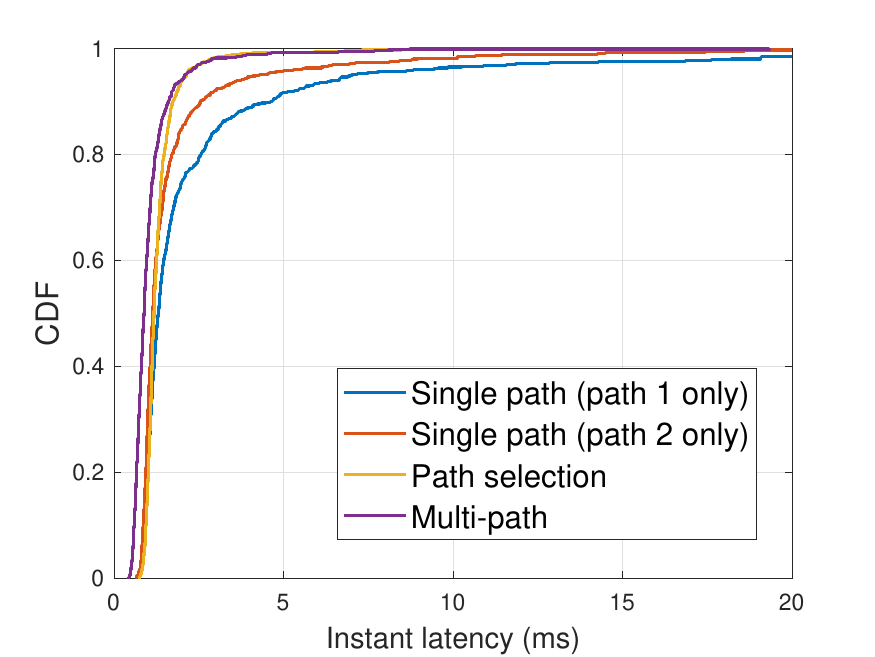}}
\label{fig:2a}
\end{minipage}
\begin{minipage}[t]{0.24\linewidth}
\centering
\subfigure[Instant latency CDF of traffic Y.]{
\includegraphics[width=1.7in]{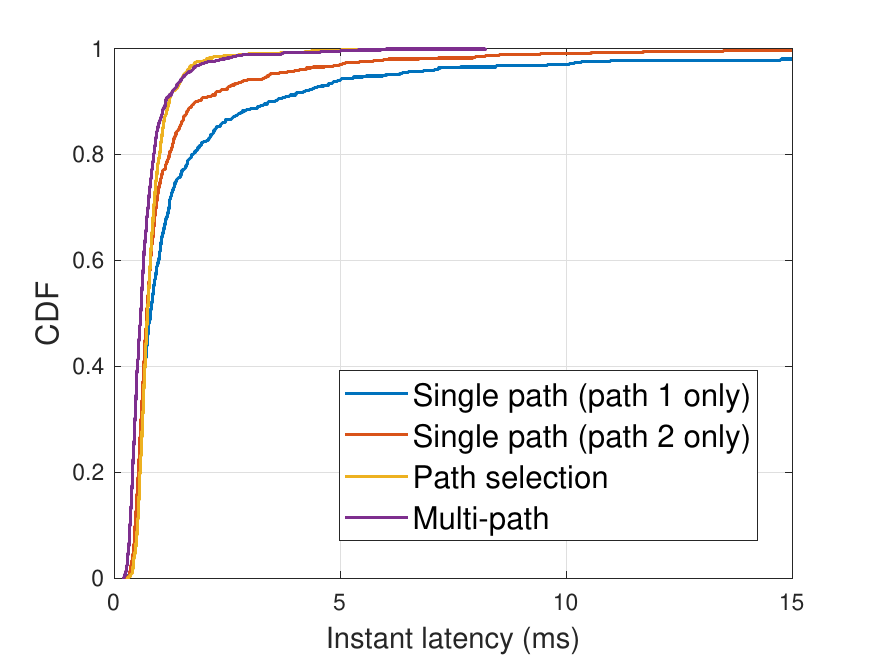}}
\label{fig:2b}
\end{minipage}%
\begin{minipage}[t]{0.24\linewidth}
\centering
\subfigure[Optimized traffic ratio.]{
\includegraphics[width=1.7in]{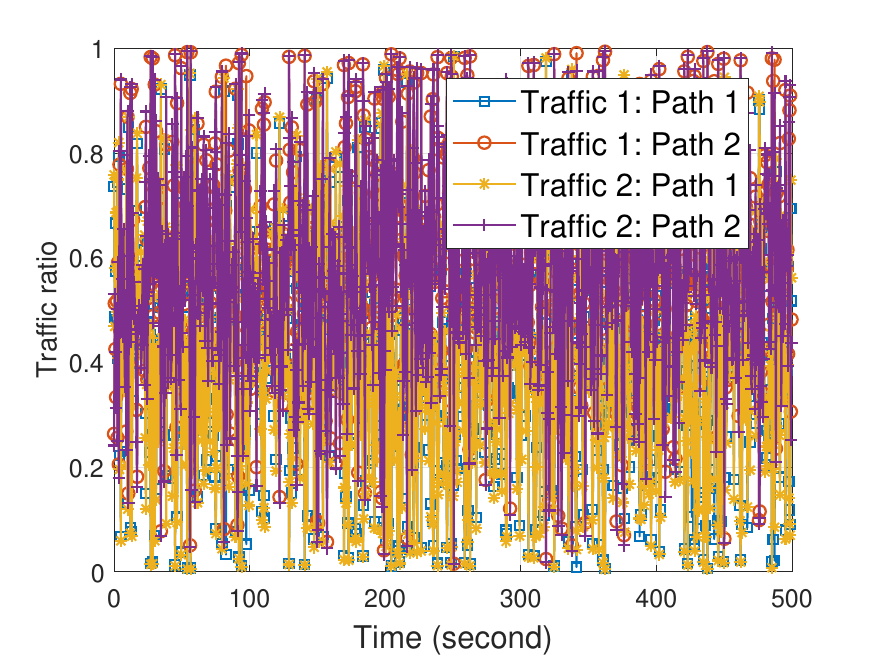}}
\label{fig:2c}
\end{minipage}
\begin{minipage}[t]{0.24\linewidth}
\centering
\subfigure[Optimized transmit power.]{
\includegraphics[width=1.7in]{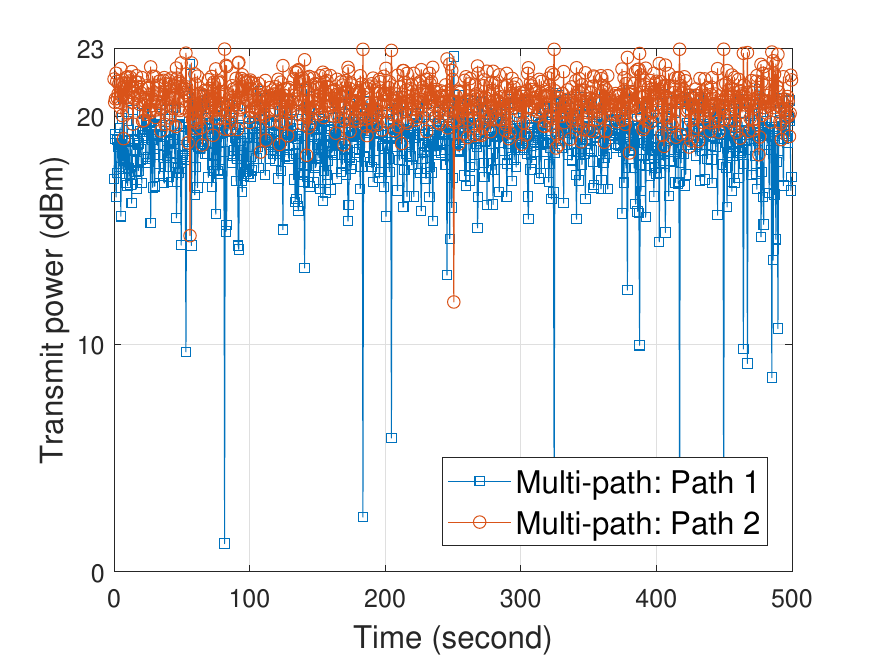}}
\label{fig:2d}
\end{minipage}
\caption{Multi-path performance under scenario 1.}
\label{fig:multiScenario1}
\end{figure*}

\begin{figure*}[t]
\begin{minipage}[t]{0.24\linewidth}
\centering
 \subfigure[Instant latency CDF of traffic X.]
{\includegraphics[width=1.7in]{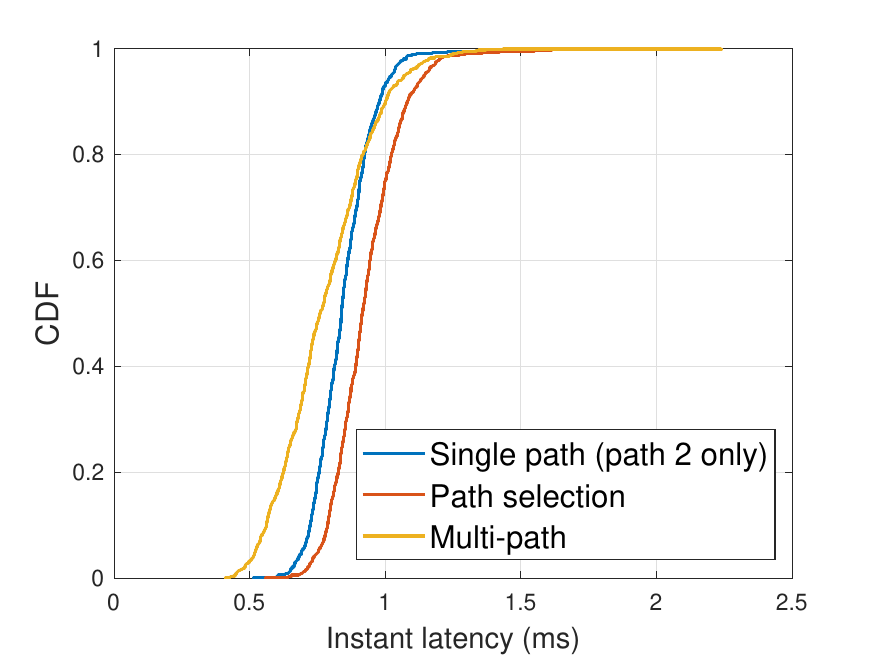}}
\label{fig:3a}
\end{minipage}
\begin{minipage}[t]{0.24\linewidth}
\centering
\subfigure[Instant latency CDF of traffic Y.]{
\includegraphics[width=1.7in]{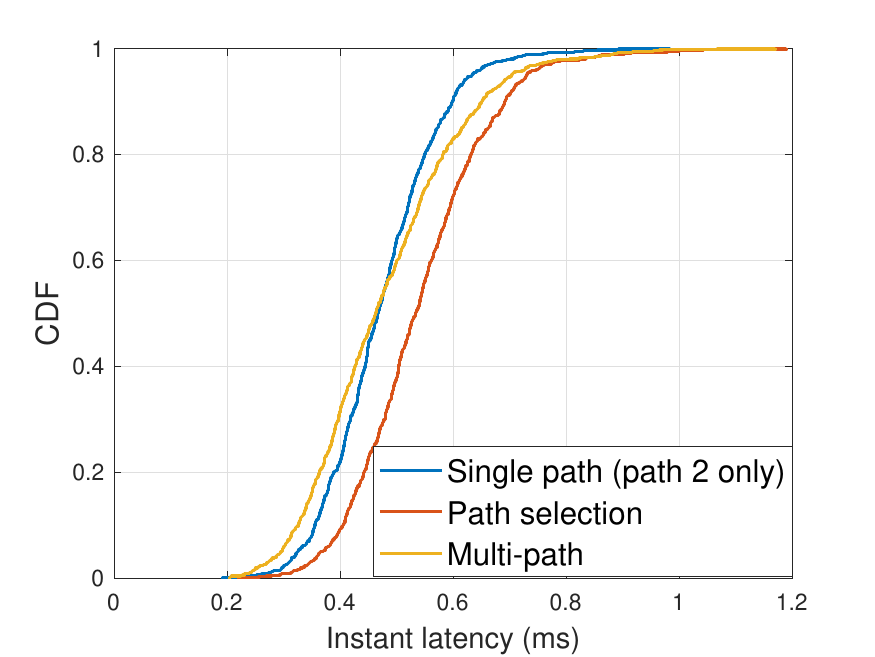}}
\label{fig:3b}
\end{minipage}%
\begin{minipage}[t]{0.24\linewidth}
\centering
\subfigure[Optimized traffic ratio.]{
\includegraphics[width=1.7in]{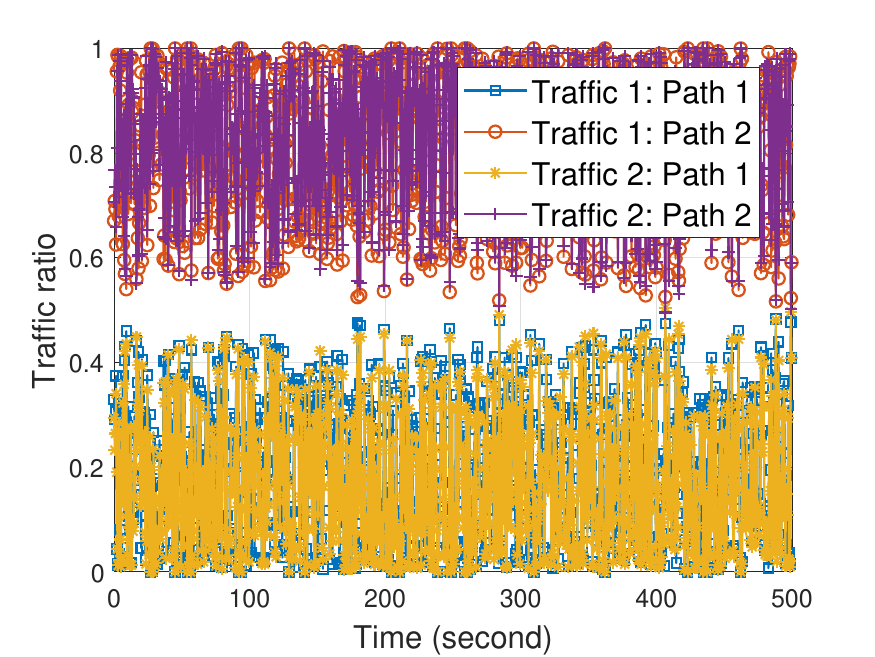}}
\label{fig:3c}
\end{minipage}
\begin{minipage}[t]{0.24\linewidth}
\centering
\subfigure[Optimized transmit power.]{
\includegraphics[width=1.7in]{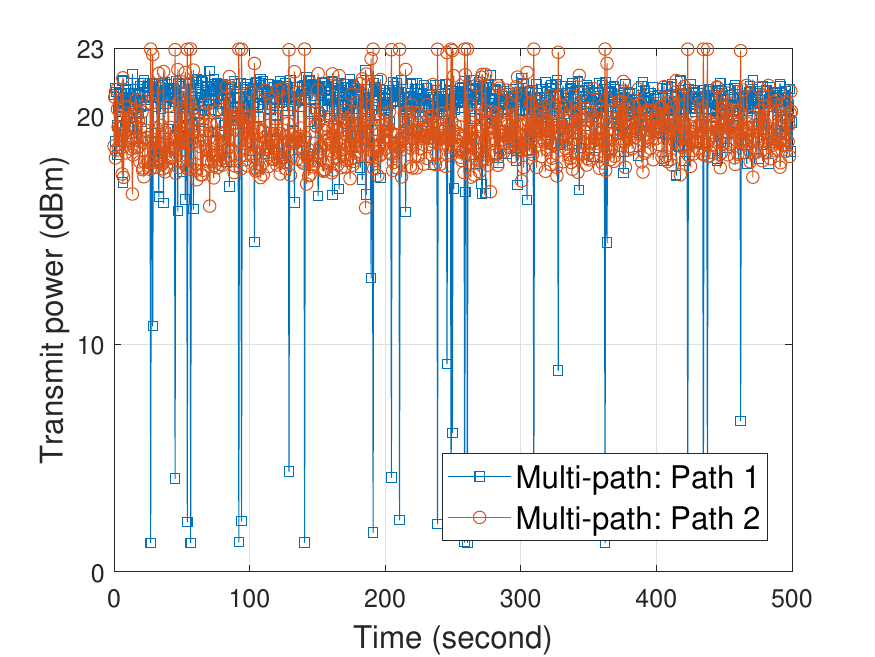}}
\label{fig:3d}
\end{minipage}
\caption{Multi-path performance under scenario 2.}
\label{fig:multiScenario2}
\end{figure*}

\begin{table}[h]
 \centering
 \caption{\small{Main simulation parameters.} 
}\label{tab:sim_para}
\resizebox{.4\textwidth}{!}{\begin{tabular}{ |c|c|c|c|c|c|c| } 
\hline
\textbf{Parameter} & \textbf{Value} \\
\hline
Packet size, $M_i$ (bytes) &  Traffic X/Y: 100/300 \\
\hline
Average packet arrival rate (pkts/s)  & Traffic X/Y: 200/50 \\
\hline
Average number of queuing packets, $n_i^q$  & Traffic X/Y: 10/5 \\
\hline
Latency constraint, $T_i^{max}$ (second) & \vtop{\hbox{\strut Traffic X/Y: 0.9/0.85}} \\
\hline
Noise PSD, $N0$ (dBm/Hz) & -174  \\
\hline
Center frequency band (GHz) & 2.6  \\
\hline
Total transmit power, $P_{tot}$ (dBm)& 23 \\
\hline
Total bandwidth (MHz)& \{10, 50, 100\}\\
\hline
Pathloss (dB) & Uma NLOS [5] \\
\hline
Shadowing, $\sigma$ (dB) & 7.8 \\
\hline
GBR range on path 1, $c_i^1$ (Mb/s) & Traffic X/Y: [100, 140]/[200, 220] \\
\hline
GBR range on path 2, $c_i^2$ (Mb/s) & Traffic X/Y: [110, 130]/[180, 200] \\
\hline
Speed of UE (m/s) & 1 \\
\hline
Time interval duration, $t_s$ (second) & 0.5 \\
\hline
Simulation time (second) & 500 \\
\hline
CPU & Intel Core i7 CPU at 2.6GHz \\
\hline
\end{tabular}}
\end{table}


The main simulation setup is summarized in Table \ref{tab:sim_para}, where the used two traffic types have different characteristics. The proposed solution is solved by using Matlab CVX. 
For a fair comparison between the proposed multi-path solution and the baselines, UE assigns the whole transmit power $P_{tot}$ to the selected path under the path-selection solution and the fixed path under the single-path solution. Meanwhile, BS1 and BS2 are assumed to have equal bandwidth (i.e., $B_1 = B_2$), and each BS has the half of total bandwidth under the multi-path solution and the path-selection solution. By contrast, the single BS utilizes the total bandwidth for the fixed path under the single-path solution. The GBR in each time interval under each solution keeps the same.

\subsection{Multi-path performance under scenario 1}
We first provide an example (Total bandwidth = 100 MHz) to compare the performance of the proposed multi-path solution with baselines under scenario 1. In this scenario, the initial UE’s proximity to each BS is set to 250 meters. 
\begin{table}[ht]
 \centering
 \caption{\small{Average Latency (ms) under scenario 1.} 
}\label{tab:aveLatencyScenario1}
\resizebox{.4\textwidth}{!}{\begin{tabular}{|lllll|}
\hline
\multicolumn{1}{|l|}{\vtop{\hbox{\strut Traffic}  \hbox{\strut  \; type}}} & \multicolumn{1}{l|}{Multi-path} & \multicolumn{1}{l|}{\vtop{\hbox{\strut Single-path} \hbox{\strut(path 1 only)}}} & \multicolumn{1}{l|}{\vtop{\hbox{\strut Single-path} \hbox{\strut (path 2 only)}}} & Path-selection \\ \hline
\multicolumn{1}{|l|}{\quad X}    & \multicolumn{1}{l|}{\quad 1.0745}     & \multicolumn{1}{l|}{\quad 2.6222}                    & \multicolumn{1}{l|}{\quad 1.7679}                    & \quad 1.3406         \\ \hline
\multicolumn{1}{|l|}{\quad Y}    & \multicolumn{1}{l|}{\quad 0.7320}     & \multicolumn{1}{l|}{\quad 1.8530}                    & \multicolumn{1}{l|}{\quad 1.2010}                    & \quad 0.8550         \\ \hline
\end{tabular}}
\end{table}

Table \ref{tab:aveLatencyScenario1} shows the average latency of two traffics among all solutions under scenario 1. Under this simulation setup, multi-path solution performs the best, while the single-path solutions perform the worst in terms of the average latency of both traffics. Fig. \ref{fig:multiScenario1} provides more performance details of the proposed multi-path solution. Fig. \ref{fig:multiScenario1} (a) and (b) show the instant latency cumulative distribution function (CDF) of traffic type X and Y, respectively. We can observe that the single-path solutions have long tails, which indicates that the UE suffers from extraordinarily higher latency compared with the multi-path and the path-selection. This is because the multi-path and path-selection always can leverage a higher quality path between the two paths while the single-path solution sticks with the only path for data transmission. Once the quality of the fixed path in a time interval is really bad, the instant latency of that time interval becomes extremely high. As a result, the average latency under the single-path increases. Note that the path-selection solution can be considered as a special case of the multi-path solution, i.e., the total transmit power and the total traffic ratio are allocated to the better path. Therefore, the optimal solution of the multi-path solution always performs better than the path-selection in terms of the average latency and instant latency.

Fig. \ref{fig:multiScenario1} (c) and (d) show the optimization decision results under scenario 1. Under the previous simulation setup, the traffic ratios allocated to two paths of both traffics are almost overlapping because the wireless link quality of the two paths is similar due to the approximate pathloss effects. Meanwhile, the average transmit power allocated to Path 2 is subtly higher than that allocated to Path 1. According to Table. \ref{tab:aveLatencyScenario1}, the average latency of Path 1 only is higher than that of Path 2 only, indicating that the wireless link quality of Path 2 is generally better than that of Path 1. Hence UE is prone to allocate more transmit power to Path 2 in the proposed multi-path solution.

\subsection{Multi-path performance under scenario 2}
We further study an example (Total bandwidth = 100 MHz) under scenario 2. In such a scenario, the initial UE’s proximity to BS1 and BS2 is 475 and 25 meters, respectively.

\begin{table}[h]
 \centering
 \caption{\small{Average latency (ms) under scenario 2.} 
}\label{tab:aveLatencyScenario2}
\resizebox{.4\textwidth}{!}{\begin{tabular}{|lllll|}
\hline
\multicolumn{1}{|l|}{\vtop{\hbox{\strut Traffic}  \hbox{\strut  \; type}}} & \multicolumn{1}{l|}{Multi-path} & \multicolumn{1}{l|}{\vtop{\hbox{\strut Single-path} \hbox{\strut(path 1 only)}}} & \multicolumn{1}{l|}{\vtop{\hbox{\strut Single-path} \hbox{\strut (path 2 only)}}} & Path-selection \\ \hline
\multicolumn{1}{|l|}{\quad X}    & \multicolumn{1}{l|}{\quad 0.7778}     & \multicolumn{1}{l|}{\quad 10.5526}                    & \multicolumn{1}{l|}{\quad 0.8460}                    & \quad 0.9297         \\ \hline
\multicolumn{1}{|l|}{\quad Y}    & \multicolumn{1}{l|}{\quad 0.4787}     & \multicolumn{1}{l|}{\quad 8.1722}                    & \multicolumn{1}{l|}{\quad 0.4738}                    & \quad 0.5423         \\ \hline
\end{tabular}}
\end{table}
Table \ref{tab:aveLatencyScenario2} shows the average latency of two traffics among all solutions under scenario 2. Under this simulation setup, the multi-path solution performs close to the single-path solution (i.e., path 2 only), while the path-selection solution performs the worst. As is shown in Fig \ref{fig:multiScenario2} (a) and (b), the instant latency CDF of the multi-path solution has a shorter tail for traffic type X while a longer tail for traffic type Y compared with the single-path solution. As Fig. \ref{fig:multiScenario2} (c) and (d) show, first, the traffic ratios allocated to two paths of both traffics under scenario 2 are such that the traffic ratio allocated to Path 2 is close to 1 while the ratio to Path 1 is close to 0 in most time. This is because the wireless link quality of Path 2 becomes much better than that of Path 1 due to less pathloss effects. Second, since UE is quite close to BS2, UE is prone to allocate more transmit power to Path 2 in the proposed multi-path solution.

\begin{figure}[h]
\begin{minipage}[t]{0.48\linewidth}
\centering
 \subfigure[Scenario 1.]
{\includegraphics[width=1.7in]{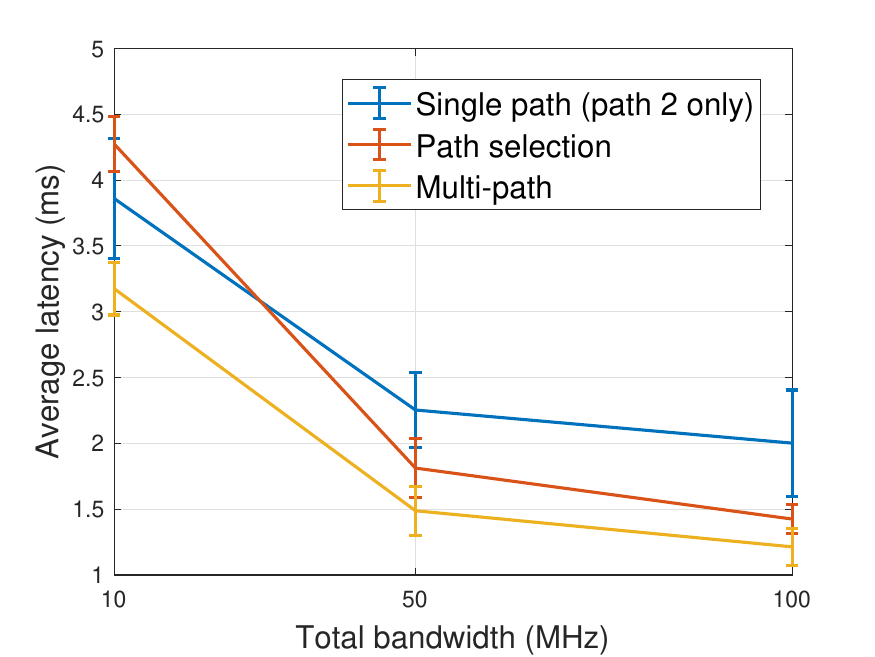}}
\label{fig:4a}
\end{minipage}
\begin{minipage}[t]{0.24\linewidth}
\centering
\subfigure[Scenario 2.]{
\includegraphics[width=1.7in]{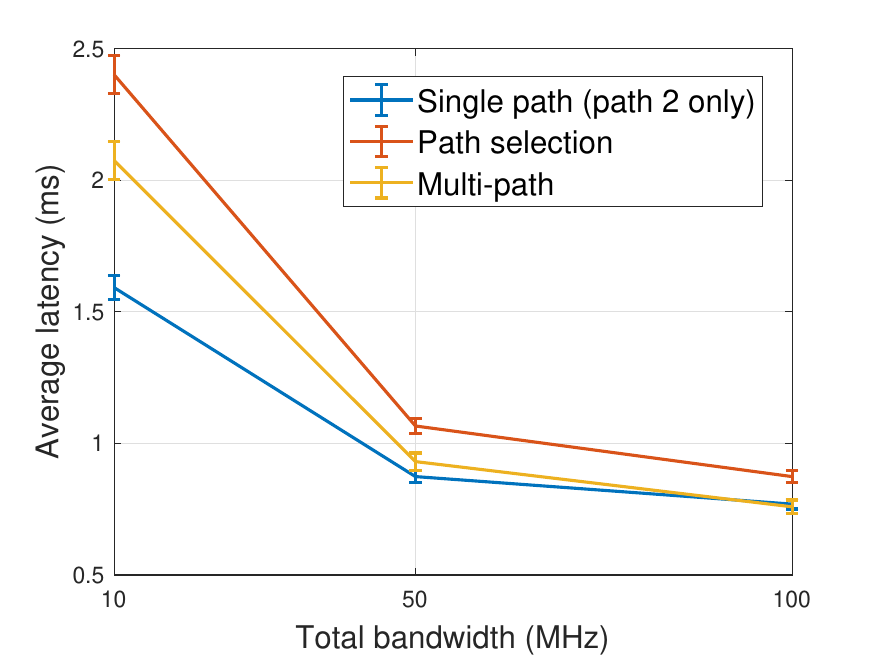}}
\label{fig:4b}
\end{minipage}%
\caption{Average latency v.s. Bandwidth.}
\label{fig:BW}
\end{figure}

\subsection{Average latency v.s. Bandwidth}
Notice that the previous examples under two scenarios only consider a high bandwidth.  We further explore the effects of bandwidth on the average latency. For simplicity, a single traffic type is considered in the simulation, where the packet size is 500 bytes and the average packet arrival rate is 50 packets/sec. Fig. \ref{fig:BW} shows the average latency of a single traffic type with different bandwidths under two scenarios. Under scenario 1, i.e., when UE walks around the center of two BSs, the proposed multi-path solution performs the best whichever the bandwidth is. By contrast, when UE walks around BS2 in most time under scenario 2, the single-path solution performs the best, especially under the lower bandwidth. However, the multi-path solution gets higher gains when increasing the bandwidth compared with the single-path solution.





\subsection{Average latency v.s. UE’s distance to the BS}
\begin{figure}[h]
\begin{minipage}[h]{0.33\linewidth}
\centering
 \subfigure[10 MHz.]
{\includegraphics[width=1.32in]{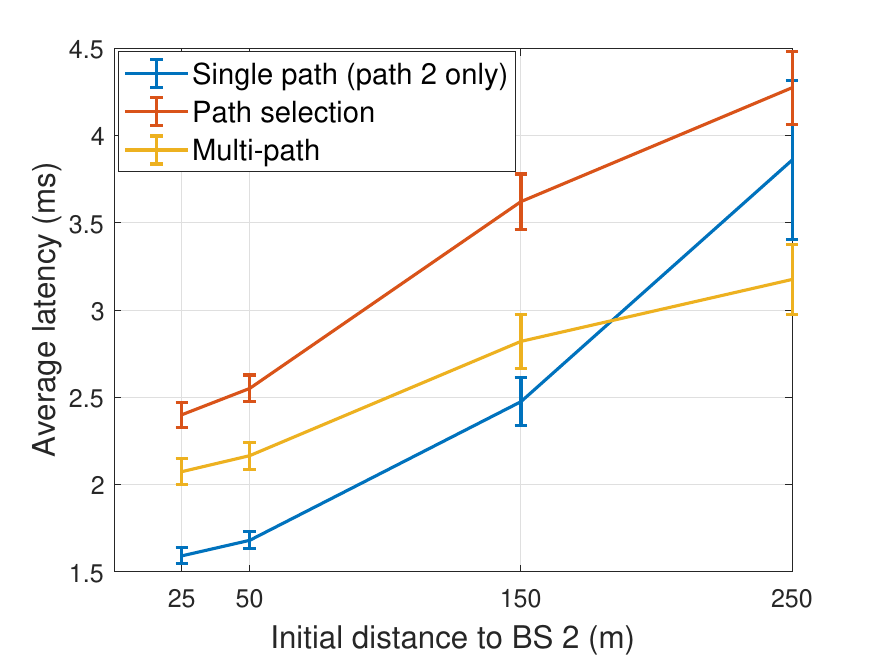}}
\label{fig:5a}
\end{minipage}
\begin{minipage}[h]{0.33\linewidth}
\centering
\subfigure[50 MHz.]{
\includegraphics[width=1.32in]{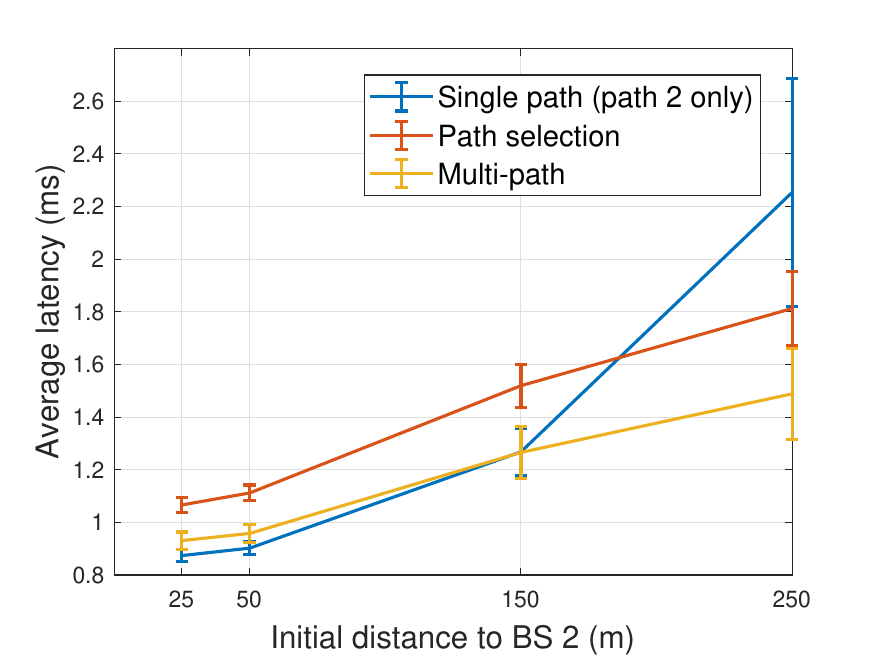}}
\label{fig:5b}
\end{minipage}%
\begin{minipage}[h]{0.33\linewidth}
\centering
\subfigure[100 MHz.]{
\includegraphics[width=1.32in]{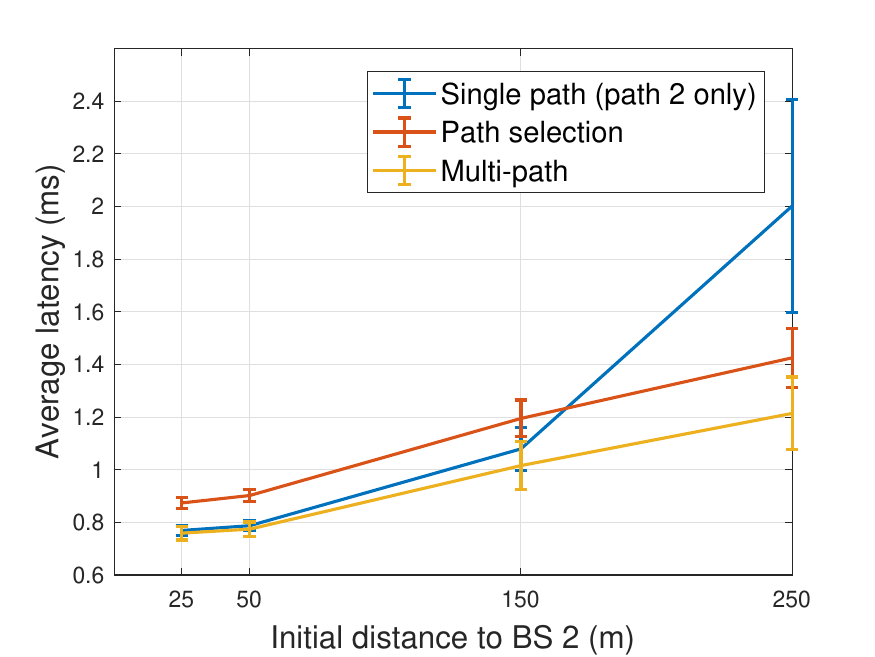}}
\label{fig:5c}
\end{minipage}
\caption{Average latency v.s. UE's distance to BS2 under different total bandwidths.}
\label{fig:distance}
\end{figure}

Afterwards, we investigate the effects of UE’s distance to the BS on the average latency of the single traffic type of which the characteristics are the same as part C. Fig. \ref{fig:distance} shows the average latency with different distances to BS2. When UE is getting away from BS2, the single-path solution gets lower gains than other solutions because the average latency under the single-path solution increases more rapidly regardless of the total bandwidth.

\subsection{Average latency v.s. Packet size}
\begin{figure}[h]
\begin{minipage}[t]{0.33\linewidth}
\centering
 \subfigure[Scenario 1: 10 MHz.]
{\includegraphics[width=1.32in]{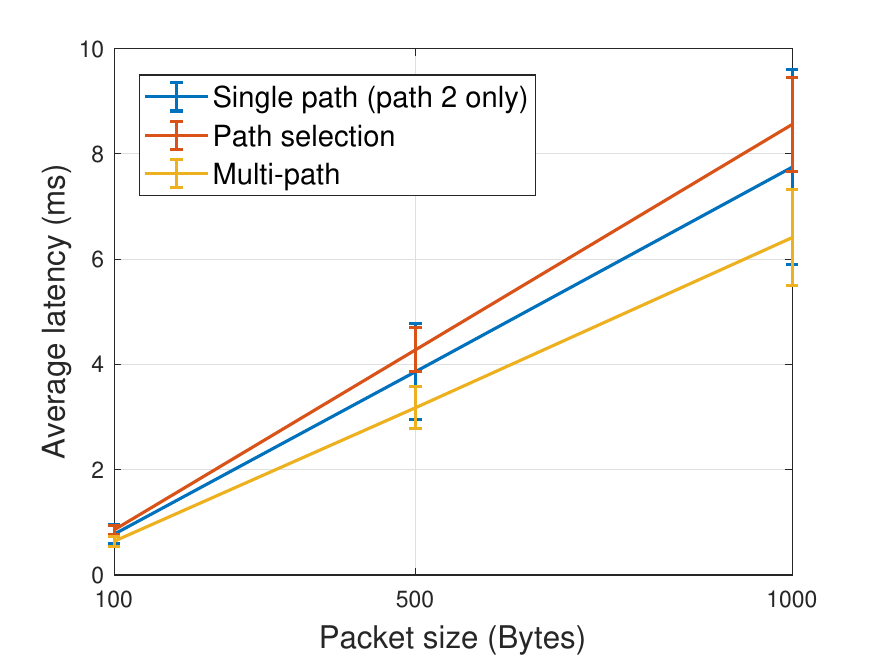}}
\label{fig:6a}
\end{minipage}
\begin{minipage}[t]{0.33\linewidth}
\centering
\subfigure[Scenario 1: 50 MHz.]{
\includegraphics[width=1.32in]{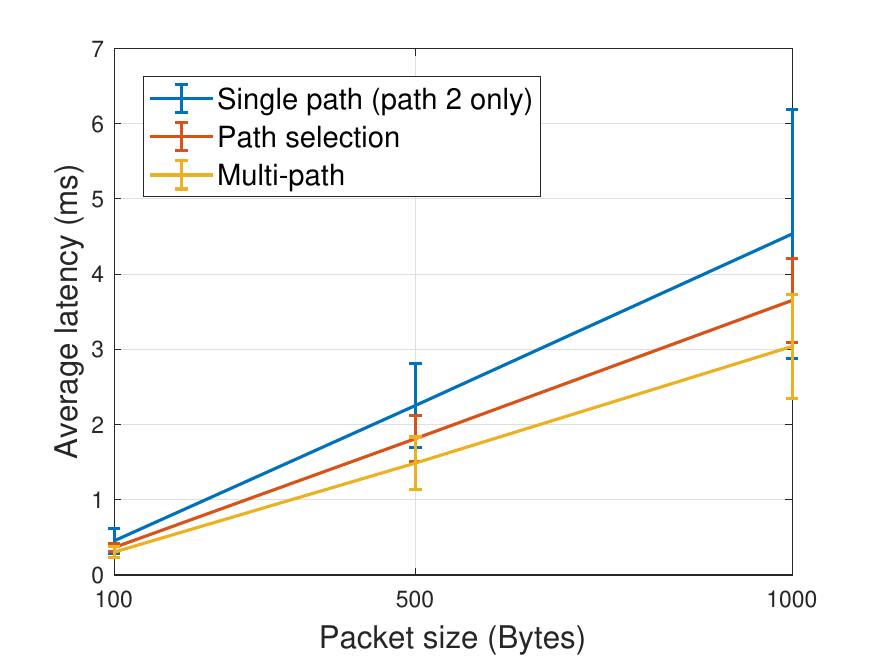}}
\label{fig:6b}
\end{minipage}%
\begin{minipage}[t]{0.33\linewidth}
\centering
\subfigure[Scenario 1: 100 MHz.]{
\includegraphics[width=1.32in]{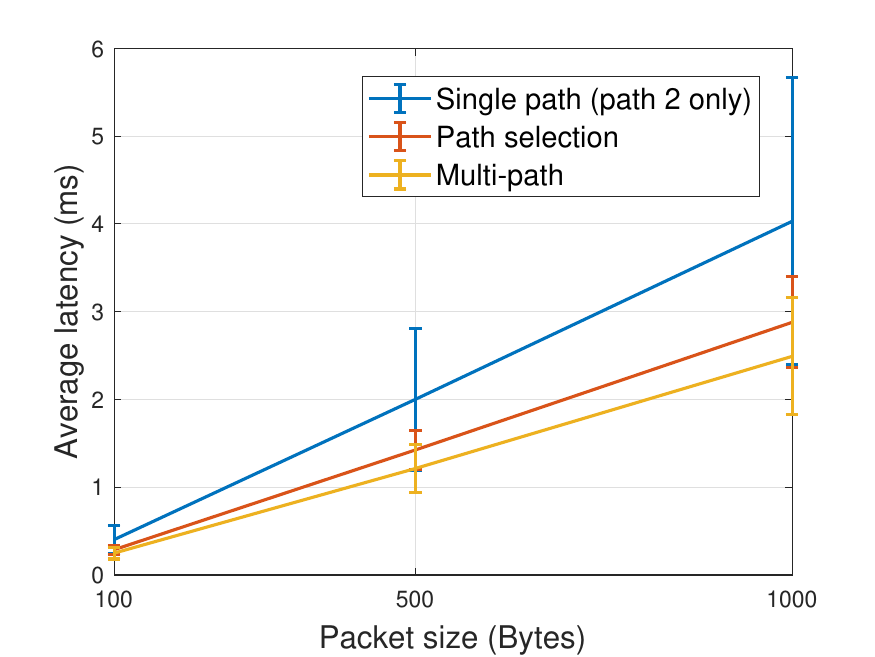}}
\label{fig:6c}
\end{minipage}

\medskip

\begin{minipage}[t]{0.33\linewidth}
\centering
 \subfigure[Scenario 2: 10 MHz.]
{\includegraphics[width=1.32in]{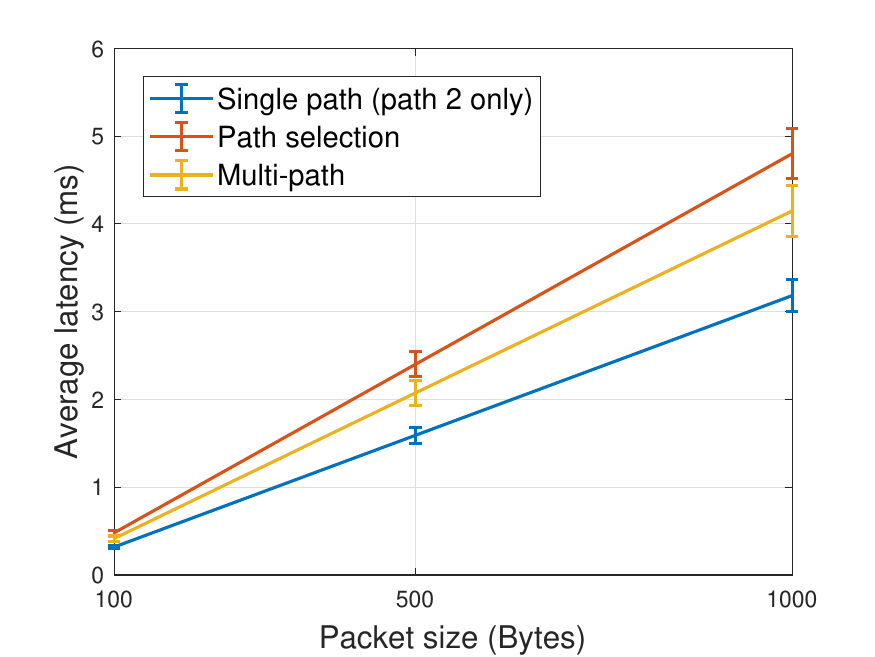}}
\label{fig:6d}
\end{minipage}
\begin{minipage}[t]{0.33\linewidth}
\centering
\subfigure[Scenario 2: 50 MHz.]{
\includegraphics[width=1.32in]{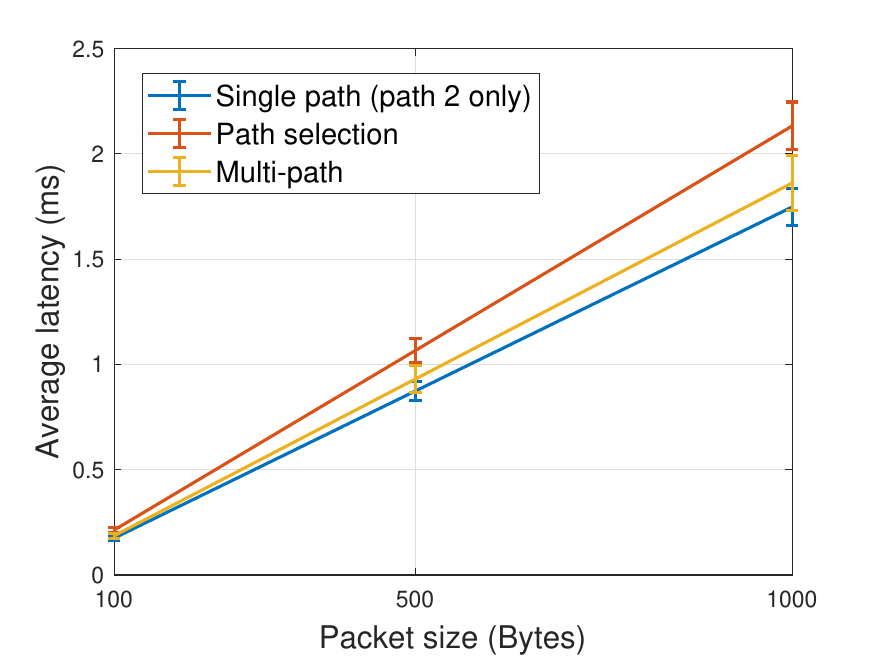}}
\label{fig:6e}
\end{minipage}%
\begin{minipage}[t]{0.33\linewidth}
\centering
\subfigure[Scenario 2: 100 MHz.]{
\includegraphics[width=1.32in]{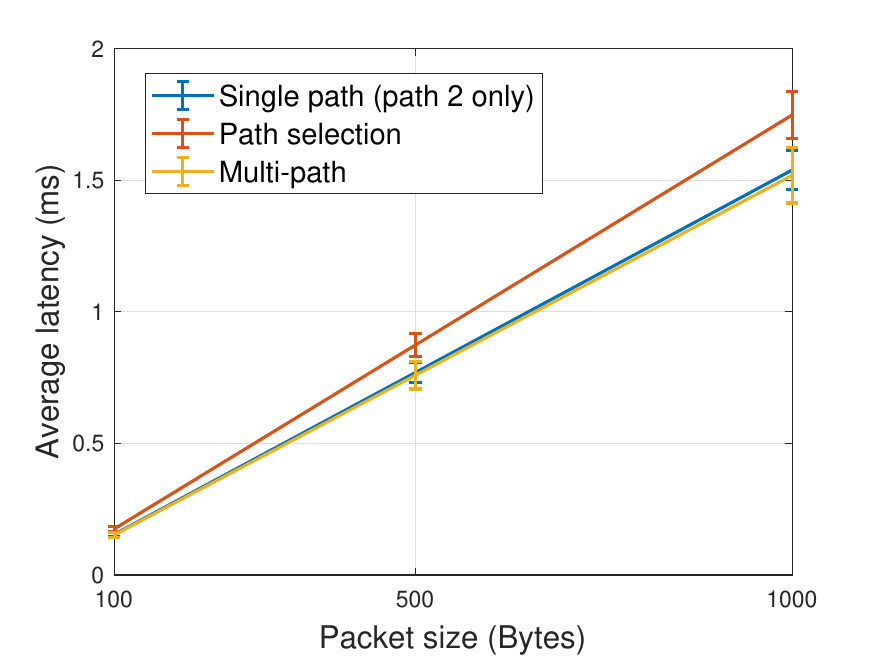}}
\label{fig:6f}
\end{minipage}

\caption{Average latency v.s. Packet size under different bandwidths.}
\label{fig:size}
\end{figure}

Finally, we study the effects of packet size on the average latency under different bandwidths, which is shown in Fig. \ref{fig:size}. Given the shown simulation setup, the average latency increases almost linearly with the increasing packet size regardless of the bandwidth. In addition, the multi-path solution performs the best when UE walks around the center of two BSs while the single-path generally performs the best when UE walks around one BS, especially under the lower bandwidth.

\section{Conclusion}
\label{sec:con}
In this paper, we proposed an E2E multi-path solution that minimizes the UE’s UL latency of multiple traffic types through the joint optimization of the traffic ratio of each traffic type on each path and the associated transmit power. We also compared the performance of the proposed E2E multi-path solution with that of the proposed baselines in terms of the instant latency in each time interval and the average latency. Particularly, given the described simulation setup, our key observations are 
\begin{itemize}
\item Multi-path solution performs the best in the situation in which UE is constantly moving between two BSs such that the pathloss effects of two paths are approximate.
\item Single-path solution usually performs the best in the situation in which UE is constantly moving around one BS such that the pathloss of the closer BS is much lower than that of the other.
\item Increasing the bandwidth makes the multi-path solution get higher gain than the single-path solution in the situation in which UE is constantly moving around one BS.
\end{itemize}

In our future work, we will take into consideration of similar optimization for downlink (DL) latency. In such a case, the queuing delay cannot be negligible. 


\bibliographystyle{IEEEtran}
\bibliography{ref}

\begin{thebibliography}{10}
\providecommand{\url}[1]{#1}
\csname url@samestyle\endcsname
\providecommand{\newblock}{\relax}
\providecommand{\bibinfo}[2]{#2}
\providecommand{\BIBentrySTDinterwordspacing}{\spaceskip=0pt\relax}
\providecommand{\BIBentryALTinterwordstretchfactor}{4}
\providecommand{\BIBentryALTinterwordspacing}{\spaceskip=\fontdimen2\font plus
\BIBentryALTinterwordstretchfactor\fontdimen3\font minus
  \fontdimen4\font\relax}
\providecommand{\BIBforeignlanguage}[2]{{%
\expandafter\ifx\csname l@#1\endcsname\relax
\typeout{** WARNING: IEEEtran.bst: No hyphenation pattern has been}%
\typeout{** loaded for the language `#1'. Using the pattern for}%
\typeout{** the default language instead.}%
\else
\language=\csname l@#1\endcsname
\fi
#2}}
\providecommand{\BIBdecl}{\relax}
\BIBdecl

\bibitem{3gpp.23.725}
3GPP, ``{Study on enhancement of Ultra-Reliable Low-Latency Communication
  (URLLC) support in the 5G Core network (5GC)},'' The 3rd Generation
  Partnership Project, Tech. Rep. {TR}23.725, June 2019.

\bibitem{3gpp.23.501}
------, ``{System architecture for the 5G System (5GS)},'' The 3rd Generation
  Partnership Project, Tech. Rep. {TS}23.501, Sept 2022.

\bibitem{3gpp.23.700}
------, ``{Study on access traffic steering, switching and splitting support in
  the 5G system architecture},'' The 3rd Generation Partnership Project, Tech.
  Rep. {TR}23.700, Jul 2020.

\bibitem{3gpp.22.841}
------, ``{Study on Upper layer traffic steer, switch and split over dual 3GPP
  access},'' The 3rd Generation Partnership Project, Tech. Rep. {TR}22.841, Aug
  2022.

\bibitem{simon2020atsc}
M.~Simon, E.~Kofi, L.~Libin, and M.~Aitken, ``{ATSC} 3.0 broadcast {5G} unicast
  heterogeneous network converged services starting release 16,'' \emph{IEEE
  Transactions on Broadcasting}, vol.~66, no.~2, pp. 449--458, 2020.

\bibitem{ko2018joint}
H.~Ko, J.~Lee, and S.~Pack, ``Joint optimization of channel selection and frame
  scheduling for coexistence of {LTE} and {WLAN},'' \emph{IEEE Transactions on
  Vehicular Technology}, vol.~67, no.~7, pp. 6481--6491, 2018.

\bibitem{singh2016proportional}
S.~Singh, M.~Geraseminko, S.-p. Yeh, N.~Himayat, and S.~Talwar, ``Proportional
  fair traffic splitting and aggregation in heterogeneous wireless networks,''
  \emph{IEEE Communications Letters}, vol.~20, no.~5, pp. 1010--1013, 2016.

\bibitem{ba2022multiservice}
X.~Ba, L.~Jin, Z.~Li, J.~Du, and S.~Li, ``Multiservice-based traffic scheduling
  for {5G} access traffic steering, switching and splitting,'' \emph{Sensors},
  vol.~22, no.~9, p. 3285, 2022.

\bibitem{hurtig2018low}
P.~Hurtig, K.-J. Grinnemo, A.~Brunstrom, S.~Ferlin, {\"O}.~Alay, and N.~Kuhn,
  ``Low-latency scheduling in {MPTCP},'' \emph{IEEE/ACM Transactions on
  Networking}, vol.~27, no.~1, pp. 302--315, 2018.

\bibitem{yin2022routing}
H.~Yin, S.~Roy, and L.~Cao, ``Routing and resource allocation for {IAB}
  multi-hop network in {5G} advanced,'' \emph{IEEE Transactions on
  Communications}, 2022.

\bibitem{yin2021scheduling}
H.~Yin, L.~Cao, and X.~Deng, ``Scheduling and resource allocation for multi-hop
  urllc network in {5G} sidelink,'' in \emph{2021 IEEE 94th Vehicular
  Technology Conference (VTC2021-Fall)}.\hskip 1em plus 0.5em minus 0.4em\relax
  IEEE, 2021, pp. 1--7.

\bibitem{zhao2019joint}
G.~Zhao, Y.~Li, C.~Xu, Z.~Han, Y.~Xing, and S.~Yu, ``Joint power control and
  channel allocation for interference mitigation based on reinforcement
  learning,'' \emph{IEEE Access}, vol.~7, pp. 177\,254--177\,265, 2019.

\bibitem{kaippallimalil20205g}
J.~Kaippallimalil and A.~Xiang, ``{5G} system architecture,'' in \emph{5G
  System Design}.\hskip 1em plus 0.5em minus 0.4em\relax Springer, 2020, pp.
  273--298.

\end{thebibliography}
\end{document}